\documentclass[10pt,letterpaper]{article}
\usepackage{opex3}

\newcommand{\G}{\mathcal{G}}
\usepackage{geometry,color,mathptmx,courier,helvet}
\usepackage{graphicx}
\usepackage{amssymb}
\usepackage{amsmath}
\usepackage{epstopdf}
\usepackage{cite}

\renewenvironment{abstract}
{\vskip1pc\noindent\begin{center} \begin{minipage}{.8\textwidth} {\bf Abstract: \ } }
{ \\ \vskip-.5pc \noindent \small \copyright \, \number 2014 \hskip.05in
   Optical Society of America \\ \hfil \end{minipage}\end{center}\normalsize\vskip-1.5pc}%
	
\begin{document}

\title{Quantum non-Gaussianity of frequency up-converted single photons}

\author{Christoph~Baune,$^1$ Axel~Sch\"onbeck,$^1$ Aiko~Samblowski,$^1$ Jarom\'ir~Fiur\'a\v{s}ek,$^2$ and Roman~Schnabel$^{1,\ast}$}
\address{
$^1$Institut f\"ur Gravitationsphysik, Leibniz Universit\"at Hannover and Max-Planck-Institut f\"ur Gravitationsphysik (Albert-Einstein-Institut), Callinstrasse 38, 30167 Hannover, Germany\\
$^2$Department of Optics, Palack\'y University, 17. listupadu 12, 77146 Olomouc, Czech Republic}

\email{$^\ast$\,roman.schnabel@aei.mpg.de} 



\begin{abstract}Nonclassical states of light are an important resource in today's quantum communication and metrology protocols. 
Quantum up-conversion of nonclassical states is a promising approach to overcome frequency differences between disparate subsystems within a quantum information network. 
Here, we present the generation of heralded narrowband single photons at 1550\,nm via cavity enhanced spontaneous parametric down-conversion (SPDC) and their subsequent up-conversion to 532\,nm. 
%
Quantum non-Gaussianity (QNG), which is an important feature for applications in quantum information science, was experimentally certified for the first time in frequency up-converted states.
\end{abstract}


\section{Introduction}
Quantum non-Gaussian states of light cannot be expressed as mixtures of Gaussian states and form a subclass of nonclassical states.
Important quantum information tasks cannot be performed with Gaussian states and Gaussian operations only, as for example stated in the no-go theorems for entanglement distillation \cite{Eisert2002,Giedke2002,Fiurasek2002}, quantum error correction \cite{Niset2009} or quantum computing \cite{Ohlinger2010}.
Non-Gaussian states are therefore an important resource in modern quantum information processing. 
Filip and Mi\v{s}ta proposed a practical criterion to verify QNG characteristics of specific quantum states of light \cite{Filip2011}, which was used to analyse single photons from SPDC \cite{Jezek2011} and quantum dots \cite{Predojevic2014}. 
The criterion was extended to show the QNG of noisy squeezed single photons \cite{Jezek2012} and to study its robustness \cite{Straka2014}. 
An alternative approach to witness QNG is demonstrated in \cite{Genoni2013}.

Quantum up-conversion is a versatile tool in today's quantum optics experiments. 
It has the potential to overcome frequency differences between disparate subsystems within a quantum information network. 
Today, the generation and fiber-based transmission of nonclassical states of light are most efficient at near infra-red wavelengths, for instance at the telecommunication wavelength of 1550\,nm  \cite{Marcikic2003}. 
Future quantum memories based on trapped atoms \cite{Boozer2007}, ions \cite{Olmschenk2009} or atomic ensembles \cite{Duan2001} will potentially use shorter wavelengths, up to the visible spectrum. 
Furthermore, the up-conversion into the visible regime enables the efficient detection of infrared single photons with commercially available, low-noise and easy-to-handle silicon avalanche photodetectors \cite{Hadfield2009}. 

To make quantum up-conversion a useful tool in quantum networks the photon statistics and in particular the nonclassical properties of the quantum states such as QNG have to be preserved.
In their pioneering work, Huang and Kumar demonstrated quantum frequency conversion of a bright beam maintaining its nonclassical intensity correlations via second harmonic generation \cite{Huang1992}. 
Recently, the up-conversion of squeezed vacuum states of light was shown \cite{Vollmer2014}.
Quantum up-conversion can also be realized with high efficiency at the single photon level via sum-frequency generation with a strongly attenuated coherent signal field \cite{Albota2004,Langrock2005,Pan2006}. 
In addition, nonclassicality of up-converted single photons from quantum dots has been demonstrated by measuring the second-order correlation function $g^{(2)}(0)$ \cite{Rakher2010}. 
However, quantum non-Gaussianity has not been certified for up-converted single photons yet. 
Here, we demonstrate for the first time QNG of frequency up-converted states. 

\section{Theoretical background}
\subsection{Cavity enhanced SPDC}\label{sec:theo}
\begin{figure}[h]
	\centering
\includegraphics[width=9cm]{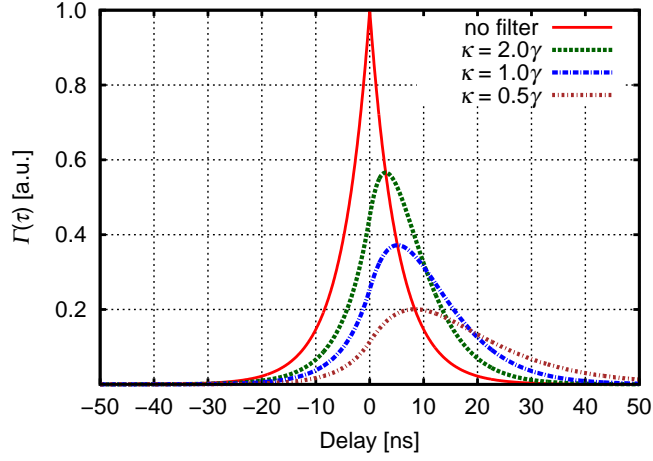}
	\caption{Calculated time shift between detection events of trigger ($\hat{a}_+$) and signal ($\hat{a}_-$) photons in our setup and the effect of extra-filtering of the signal mode. The narrower the linewidth $\kappa$ of the extra filter in the signal path, the more the correlations between trigger and signal events are smoothed out. Parameters are $\gamma=\pi\cdot$31\,MHz, $\varepsilon=0.10\gamma$\,.}
	\label{fig:corrFilNew}
\end{figure}

Spontaneous parametric down-conversion (SPDC) is a widely used technique to produce correlated photon pairs. 
It generally utilizes a non-linear crystal, which is pumped with a laser field. 
In our experiment, this crystal is embedded in an optical cavity. 
If such a device is operated above its oscillation threshold to produce bright fields in well defined spatial output modes, it is commonly known as an optical parametric oscillator (OPO). 
Far below the oscillation threshold, in a SPDC process photon pairs are generated to a very good approximation.
The Heisenberg equations of motion for the cavity modes of interest are \cite{Drummond1990}
\begin{equation}
\frac{d\hat{a}_\pm(t)}{dt}=-\frac{i}{\hslash}\left[\hat{a}_\pm(t),\textbf{H}_{\rm sys}\right]-\gamma\hat{a}_\pm(t)+\sqrt{2\gamma}\hat{a}_{\rm \pm,in}(t)\,,
\end{equation}
where $\hat{a}_{\pm}$ ($\hat{a}_{\pm\rm, in}$) are the (input) mode operators for the upper (+) and lower (-) sideband, $\gamma$ is the mean decay rate of the cavity, and the system Hamiltonian is given by
\begin{equation}\begin{split}
\textbf{H}_{\rm sys}=&\hslash\omega_+\hat{a}^\dagger_+\hat{a}_++\hslash\omega_-\hat{a}^\dagger_-\hat{a}_-\\
&+i\hslash\left(\varepsilon{\rm e}^{-i\omega_{\rm p}t}\hat{a}^\dagger_+\hat{a}^\dagger_--\varepsilon^*{\rm e}^{i\omega_{\rm p}t}\hat{a}_+\hat{a}_-\right)\,.
\end{split}\end{equation}
The gain parameter $\varepsilon=\gamma\sqrt{P/P_{\rm th}}$, with $P_{\rm th}$ being the threshold pump power for parametric oscillation, is proportional to the mean pump field amplitude $\sqrt{P}$.
The set of differential equations can be solved with straightforward methods. 
The temporal correlation function of the two output modes is \cite{Drummond1990}
\begin{equation}\begin{split}
\Gamma(\tau)=&\langle \hat{a}^\dagger_{+,\rm out}(t)\hat{a}^\dagger_{-,\rm out}(t+\tau)\hat{a}_{-,\rm out}(t+\tau)\hat{a}_{+,\rm out}(t)\rangle\\
=&\left[\frac{\varepsilon\gamma}{2}\left(\frac{1}{\lambda}\exp(-\lambda|\tau|)+\frac{1}{\mu}\exp(-\mu|\tau|)\right) \right]^2\,,
\end{split}\end{equation}
where $\lambda=\gamma-|\varepsilon|$, $\mu=\gamma+|\varepsilon|$.
In the experiment the modes corresponded to propagating TEM$_{00}$ modes at 810 (trigger) and 1550\,nm (signal). 
The signal mode $\hat{a}_{\rm -,out}$ experiences a stronger frequency filtering than the trigger mode because the SPDC cavity decay rate is smaller at 1550\,nm than at 810\,nm. The signal mode additionally experiences frequency filtering in the up-conversion cavity. 
We model the total filtering at 1550\,nm with the transformation 
\begin{equation}
\hat{a}_-(t)\rightarrow \hat{a}_-(t^\prime)=\int_{-\infty}^{t^\prime}dy\,\kappa\exp\big((-\kappa(t^\prime-y)\big)\hat{a}_-(y)\,,
\end{equation}
where $\kappa$ is the bandwidth of the combined filter effect.
The temporal correlations then read
\begin{equation}\label{eq:filteredMode}\begin{split}
\Gamma(\tau)=&\bigg[\frac{\gamma\varepsilon\kappa}{2}\Big(\frac{\exp(-\mu|\tau|)}{\mu(\kappa-s(\tau)\mu)}+\frac{\exp(-\lambda|\tau|)}{\lambda(\kappa-s(\tau)\lambda)}\\
&-\left(1+s(\tau)\right)\exp(-\kappa|\tau|)\frac{2\kappa^2-\lambda^2-\mu^2}{(\kappa^2-\lambda^2)(\kappa^2-\mu^2)}\Big)\bigg]^2\,,
\end{split}
\end{equation}
where $s(\tau)=+1$ for $\tau\geq 0$ and $s(\tau)=-1$ for $\tau<0$.
In Fig.~\ref{fig:corrFilNew} the effect of additional filtering of the signal mode is shown. 
The narrower the linewidth of the extra filter, the more the temporal correlations are smoothed out.

\subsection{Quantum up-conversion}
Quantum up-conversion increases the optical frequency of a mode without destroying its quantum coherences. We use sum-frequency generation, where a nonlinear crystal is pumped with a strong coherent pump field. A signal field interacts with the pump and is up-converted to a field with a frequency equal to the sum of the two initial frequencies. In 1990, Kumar showed that the quantum properties of the signal field can be fully transferred to the up-converted field \cite{Kumar1990}. 

The Hamiltonian for this process is given by \cite{Kumar1990}
\begin{equation*}
\textbf{H}_{\rm QUC}=i\hslash\zeta\left(\hat{a}_u^{}\hat{a}_f^\dagger+\hat{a}_u^\dagger\hat{a}_f^{}\right)\,,
\end{equation*}
where $\zeta$ is the coupling constant, proportional to the mean pump amplitude and the second-order susceptibility $\chi^{(2)}$. The evolution of the signal (fundamental $\hat{a}_f^{}$ and up-converted $\hat{a}_u^{}$) is 
\begin{equation*}
\hat{a}_u^{}(t)=\hat{a}_u^{}(0)\cos(\zeta t)+\hat{a}_f^{}(0)\sin(\zeta t)\,.
\end{equation*}
After the time $t_c=\pi/2\zeta$, the input field is completely converted into the output state: $\hat{a}_u^{}(t_c)=\hat{a}_f^{}(0)$.

\subsection{Quantum non-Gaussianity (QNG)} \label{sec:QNG}
The property of quantum non-Gaussianity was originally proposed by Filip and Mi\v{s}ta \cite{Filip2011}. 
Let $\G$ be the set of all Gaussian states and mixtures thereof. 
Any state $\rho$ which is not in $\G$ is defined to be quantum non-Gaussian. 
The quantum non-Gaussian states can be conveniently identified using a simple and experimentally feasible criterion based on photon number probabilities. 
For a given (measured) vacuum probability $p_0$ a maximum single-photon probability $p_1$ achievable by a Gaussian state can be calculated. 
If the measured $p_1$ is higher, then $\rho\notin\G$. 
A detailed description of how to apply this criterion can be found in \cite{Jezek2011}. 
It is also possible to define a witness of quantum non-Gaussianity, $W=p_1+ap_0-W_G(a)$, where $a<1$ is a parameter specifying the witness and $W_G(a)$ represents the maximum of $p_1+ap_0$ achievable by Gaussian states. 
Assuming Poissonian statistics of the measured coincidences, the statistical error of the witness can be determined and one can express the witness in numbers of standard deviations $\Delta W$.   
A positive witness certifies QNG of the state, even though in general it can be a mixed state with a positive Wigner function. 

\section{Experimental setup}
\begin{figure}[h]
	\centering
	\includegraphics[width=10cm]{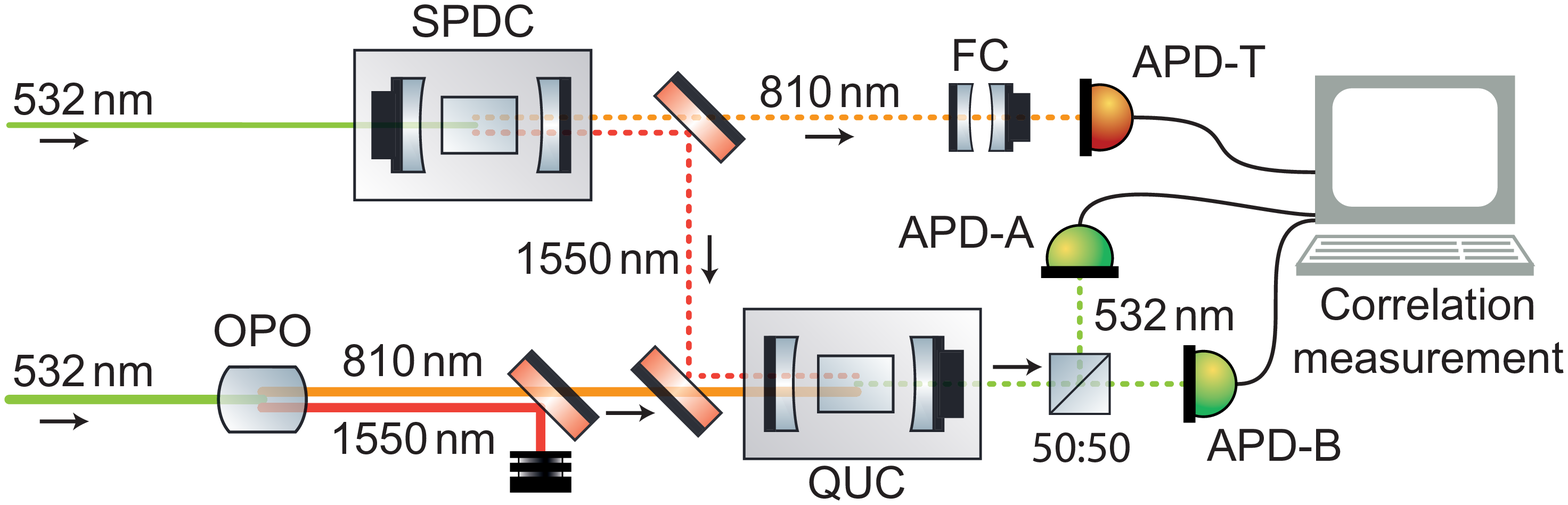}
	\caption{Schematic of the experimental setup. Two doubly resonant optical parametric oscillators are pumped above (OPO) and below (SPDC) threshold with a continuous wave 532\,nm pump field, producing bright fields and twin photons at 810\,nm and 1550\,nm. The 1550\,nm photons are up-converted to 532\,nm in the quantum up-converter (QUC) which is pumped with a strong pump field at 810\,nm and analysed in a Hanbury Brown and Twiss setup with Si-APDs (APD-A and -B). The 810\,nm photons heralding the existence of a 532\,nm photon are detected at APD-T after transmitting the filter cavity (FC).}
	\label{fig:Sipcon-Setup}
\end{figure}
The experimental setup is depicted in Fig. \ref{fig:Sipcon-Setup}. 
The main laser source was a Nd:YAG laser with 2\,W output power, that was frequency doubled to 532\,nm in a LiNbO$_3$-based second harmonic generator (not shown in the figure). 
A major fraction of the 532\,nm field pumped a monolithic non-degenerate, doubly resonant optical parametric oscillator (OPO) above threshold, providing up to 200\,mW at 810\,nm to pump the quantum up-converter (QUC). 
A smaller fraction of the 532\,nm field was used as the pump field for another nonlinear cavity (SPDC), which was operated below threshold and generated twin photons at 810\,nm and 1550\,nm in a cavity enhanced SPDC process. 
The reflectivities of the mirrors of these two cavities were AR@532\,nm, HR@810\&1550\,nm on the incoupling mirror and HR@532\,nm, 96.5\,\%@810\&1550\,nm on the outcoupling mirror. 
The linewidths of the generated modes of the SPDC cavity at 810\,nm and 1550\,nm were 35\,MHz and 19\,MHz (FWHM), respectively, determined by scanning the cavity and comparing the Airy pattern with frequency markers introduced by a phase modulation \cite{Brueckner2010}.

The QUC cavity was modified from to what was reported in \cite{Vollmer2014, Samblowski2014}. 
The new incoupling mirror had reflectivities of HR@532\,nm, 97\,\%@810\,nm and 91\,\%@1550\,nm. 
The linewidth for the signal at 1550\,nm was measured to be 68\,MHz. 
With these modifications, the up-conversion efficiency could be increased to 90.2$\,\pm\,$1.5\,\% for a pump power of 140\,mW, determined with the methods described in \cite{Samblowski2014}. 
Periodically poled potassium titanyl phosphate (PPKTP) crystals were embedded in all three cavities, actively stabilized to the phase matching temperatures.

The 810\,nm heralding photons passed a Fabry Perot filter cavity to suppress uncorrelated modes. 
This quasi-monolithic cavity consisted of two mirrors separated by 2.5\,mm, each with a reflectivity of 99\,\%@810\,nm yielding a free spectral range of 60\,GHz and a full width at half maximum of 190\,MHz (Finesse$\,\approx\,$315). 
The transmitted photons were detected with a Si-APD (\textit{Perkin Elmer} SPCM-AQRH-13).

The 532\,nm photons were split on a balanced beamsplitter and detected with Si-APDs (\textit{Laser Components} COUNT-250B). 
All APD signals were recorded with an oscilloscope (\textit{Agilent} DSO7014A). 
In each measurement we took 2000 data streams with a duration of 2\,ms each. 

\section{Results}
The signal count rates were as high as 50\,kHz, 120\,kHz and 400\,kHz, while increasing the gain parameter (i.e. the pump power) from $\varepsilon=0.10\gamma$ to $\varepsilon=0.16\gamma$ and $\varepsilon=0.28\gamma$, respectively. 
The propagation efficiency (including 90.2\,\% up-conversion efficiency) was determined to be 68\,\% by operating SPDC above threshold and comparing the 1550\,nm power with the up-converted field at 532\,nm directly in front of the APDs. The APD quantum efficiency was 60-70\,\% (according to the manufacturer's datasheet). False triggering, due to insufficient filtering of the 810\,nm mode, reduced the final detection efficiency to about 20\,\%.

\begin{figure}[h]
	\centering
		\includegraphics[width=9cm]{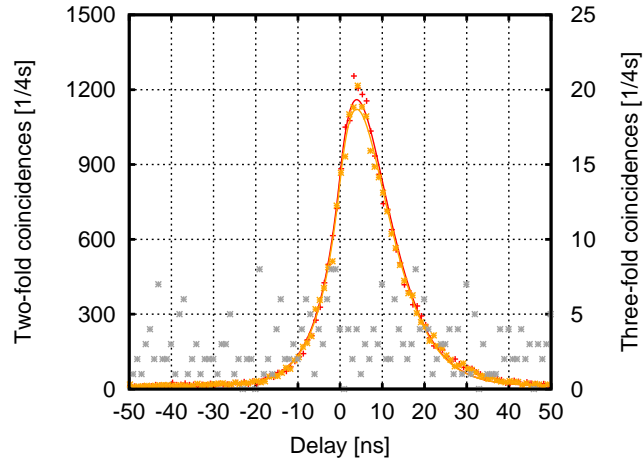}
	\caption{Histogram of the two-fold coincidence detections at APD-T and APD-A (red), and APD-T and APD-B (yellow) with theoretical curves ($\gamma=\pi\cdot31\,$MHz, $\kappa=1.4\gamma$, $\varepsilon=0.10\gamma$). The delay for the three-fold coincidences of APD-T, APD-A and APD-B (grey points, right y-axis) is defined as the time between counts at APD-A and -B given that the trigger APD-T detected a photon (within a 100\,ns time window).}
	\label{fig:histogram_low}
\end{figure}
The counting statistics of APD-A and -B triggered on APD-T are shown for $\varepsilon=0.10\gamma$ in Fig.~\ref{fig:histogram_low}. 
The temporal profile of the up-converted photons clearly shows the smoothed out exponential decay predicted by Eq.~(\ref{eq:filteredMode}). 
The theoretical curve was obtained by using a mean decay rate of $\gamma=\pi\cdot31\,$MHz and an extra filtering effect of the signal mode with $\kappa=1.4\gamma$. 
As already briefly discussed in section \ref{sec:theo}, this filtering of the signal mode is caused by asymmetric decay rates of the SPDC cavity and the transmission of the signal through the up-conversion cavity. Note that due to the relatively large linewidth of the filter cavity FC (190\,MHz), the trigger mode experiences only negligible frequency filtering.
In this graph, three-fold coincidence events do not show any significant contribution to the statistics.

\begin{figure}[h]
	\centering
		\includegraphics[width=9cm]{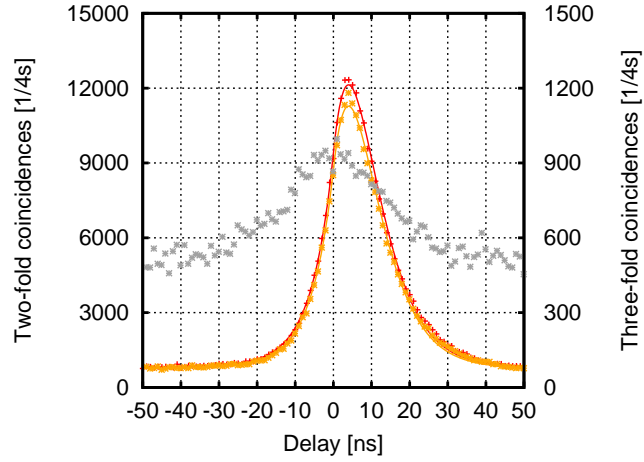}
	\caption{Histogram of the two- and threefold coincidences, as in Fig.~\ref{fig:histogram_low} with the only difference being that $\varepsilon=0.28\gamma$. The three-fold coincidences significantly contribute to the statistics.}
	\label{fig:histogram_high}
\end{figure}

When the pump power for SPDC is higher, the probability of generating more than one photon pair within the same coherence time increases. 
In Fig.~\ref{fig:histogram_high} the counting statistics are shown for a measurement with $\varepsilon=0.28\gamma$, i.e. 2.8 times the pump amplitude compared to the measurement shown in Fig.~\ref{fig:histogram_low}. The three-fold coincidences indicated by the grey circles now significantly contribute to the counting statistics. 

\begin{figure}[h]
	\centering
		\includegraphics[width=9cm]{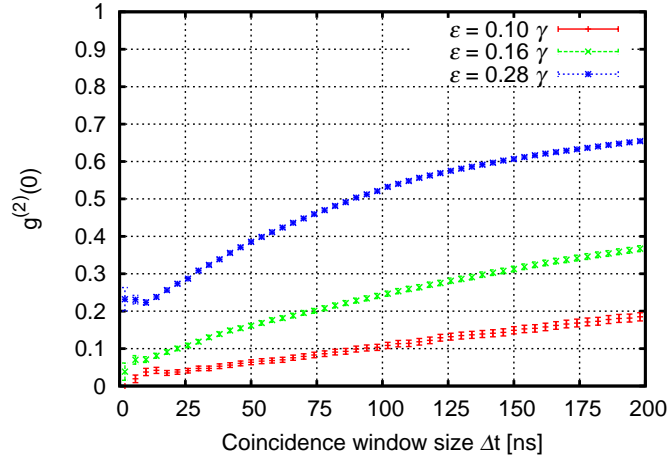}
	\caption{$g^{(2)}(0)$ values in dependence of the coincidence window $\Delta t$ and gain parameter $\varepsilon$. For all $\varepsilon$, the states show (nonclassical) subpoissonian statistics. For increasing coincidence windows more background noise is recorded and the multiphoton probability $(1-p_0-p_1)$ increases.}
	\label{fig:g2}
\end{figure}
The second order coherence function value, is calculated by $g^{(2)}(0)=2(1-p_0-p_1) / [2(1-p_0)-p_1]^2$, where the probabilities $p_0$ and $p_1$ were estimated from the measured count rates.
The two-fold ($R_{1A}$ and $R_{1B}$) and three-fold ($R_2$) coincidence rates were experimentally determined for a certain number of trigger events $R_0$. The vacuum probability is then given by
\begin{equation}
p_0=1-\frac{R_{1A}+R_{1B}+R_2}{R_0}
\end{equation}  
and the single photon probability can be bounded from below by
\begin{equation}
p_1>\frac{R_{1A}+R_{1B}}{R_0}-\frac{T^2+(1-T)^2}{2T(1-T)}\frac{R_2}{R_0}\,,
\end{equation}
where the beam splitter ratio is given by $T=R_{1A}/(R_{1A}+R_{1B})$. Further details can be found in \cite{Jezek2011}.
The measured $g^{(2)}(0)$ values are plotted against the coincidence window $\Delta t$ in Fig.~\ref{fig:g2}, for three different gain parameters $\varepsilon$. 
The coincidence window is the time interval, symmetric about the detection of a photon in APD-T, within which detections of photons in APD-A and APD-B are considered a coincidence detection.
The $g^{(2)}(0)$ reaches values smaller than 0.05 for coincidence windows less than 34\,ns, when $\varepsilon=0.10\gamma$. 
For increasing coincidence windows the $g^{(2)}(0)$ value increases as more three-fold coincidence events are registered. 
If the gain parameter $\varepsilon$ (i.e. the pump amplitude) is increased, the multiphoton probability $(1-p_0-p_1)$ increases, as does $g^{(2)}(0)$.
For very small coincidence windows ($<$5\,ns) the absolute number of coincidences is small, causing the statistical fluctuations in Fig.~\ref{fig:g2}.

\begin{figure}[h]
	\centering
		\includegraphics[width=9cm]{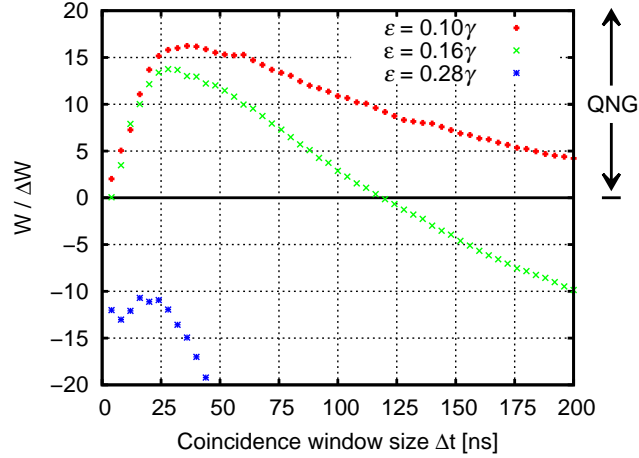}
	\caption{Quantum non-Gaussianity of the up-converted states. The witness of QNG, in numbers of standard deviations, is plotted against the size of the coincidence window and three different gain parameters $\varepsilon$. If the witness is positive, then $\rho\notin\G$. QNG could be verified with more than 16 standard deviations when $\varepsilon=0.10\gamma$ and $\Delta t$=34\,ns.}
	\label{fig:Gaussianity}
\end{figure}

The sub-Poissonian statistics indicated by a $g^{(2)}(0)$ smaller than unity are a strong signature for the nonclassicality of the states. 
An even stronger feature of nonclassicality is a negative Wigner function, the hallmark of single-photon states \cite{Lvovsky2001}. 
However, due to mixture with vacuum, the negativity of such states decreases, or even vanishes, when losses exceed 50\,\%. 
Nevertheless, it is still possible to certify the quantum non-Gaussianity of the state, a useful feature for quantum information tasks. 
The measured probabilities $p_0$ and $p_1$ are therefore used to analyze whether the up-converted states $\rho$ are quantum non-Gaussian or not. 
The result is shown in Fig.~\ref{fig:Gaussianity}. 
In the plot the witness of QNG, $W$, in numbers of standard deviations, $\Delta W$, is shown for measurements with three different gain parameters $\varepsilon$, dependent on the coincidence window size $\Delta t$. 
For each $\varepsilon$ and $\Delta t$, the parameter $a$ of the QNG witness $W$ was optimized such as to maximize $W/\Delta W$. If the witness is positive, then $\rho\notin\G$. 
For low gain $\varepsilon=0.10\gamma$, the witness of QNG was positive for all coincidence windows smaller than 300\,ns, reaching a maximum of 16 standard deviations at $\Delta t$=34\,ns. 
For higher gain, the multiphoton contribution increases, which lowers the witness. 
For $\varepsilon=0.16\gamma$ and $\Delta t\geq$120\,ns the witness is negative, indicating a state that possibly could be expressed as a convex mixture of Gaussian states. 
For even higher gain ($\varepsilon=0.28\gamma$) the witness of QNG is never positive, since as the mode approaches thermal statistics \cite{Yurke1987}, higher photon number contributions show up and background noise increases.

The maximum achievable witness can be calculated by applying the following model. 
We assume an initial two-mode squeezed vacuum state, produced by the SPDC and characterized by the pump parameter $\varepsilon$. 
Both modes propagate through lossy channels with transmittances $\eta_S$ (signal) and $\eta_T$ (trigger) that account for all imperfections of the setup, i.e. quantum efficiency of the APDs, limited up-conversion efficiencies or other optical losses. 
With these considerations, one can calculate the count rates $R_0$, $R_1$ and $R_2$ and thereby determine the probabilities $p_0$ and $p_1$. 
The maximum achievable witness $W_{\rm max}$ can then be obtained by maximizing the witness over the parameter $a$, cf. section~\ref{sec:QNG}. 
The model predicts that $W_{\rm max}$ decreases with increasing gain $\varepsilon$, which is consistent with the experimental data. 
The highest value of of $W_{\rm max}$ is therefore achieved in the weak squeezing limit ($\varepsilon\rightarrow 0$), where higher photon numbers become negligible and the single photon probability is simply given by the signal detection efficiency $\eta_S$, $p_1=\eta_S$, and $p_0=1-\eta_S$. 
In our experiment, this efficiency is bounded from below by $\eta_S=0.2$, which yields a maximum witness $W_{\rm max}=0.00486$. 
Evaluating the experimental data, the maximum witness that we could achieve was $W_{\rm exp}=0.00315$, when $\varepsilon=0.10\gamma$. 
As this reduction was most likely caused by dark counts and background noise, that were not included in the model, our experimental data are in good agreement with the theoretical predictions. 
Furthermore, the experiment was not performed exactly in the weak squeezing limit.
At least the gain had to be $\varepsilon=0.10\gamma$ to achieve sufficiently high probability of conditional generation of the single photon state and to accumulate enough data during the experiment.

\section{Discussion and conclusion}
We present the frequency up-conversion of heralded single photons from 1550 to 532\,nm.
For the first time, frequency up-converted states were shown to possess quantum non-Gaussianity and we achieved a significance of up to 16 standard deviations. 
The single photon contribution in the states with QNG certified with more than five standard deviations was up to 22.6\,\% (with $\varepsilon=0.16\gamma$, $\Delta t=80$\,ns).
The $g^{(2)}(0)$ values were smaller than 0.05 ($\varepsilon=0.10\gamma$, $\Delta t\leq34$\,ns).
Our result shows that negative Wigner functions are also within reach in up-converted states when the optical losses were minimized. 
In particular, insufficient filtering of the trigger mode, limited quantum efficiencies of the APDs and propagation losses prevented us from fully benefiting from the high up-conversion efficiency, which was determined to be $90.2\pm1.5\,\%$. 
One or two additional filter cavities with different free spectral ranges would suppress uncorrelated modes more efficiently, but a more complex setup with technically elaborate cavity locking techniques would be required. 
Anti-reflection coated fibers would minimize the propagation losses, since the coupling efficiency could be increased and an optical isolator would no longer be required. 
Nonetheless, with such improvements we expect that our setup is capable of reaching a single photon contribution of 60\,\%, mainly set by the high up-conversion efficiency ($\approx90\,\%$) and quantum efficiency of the APDs (70\,\%).

\section*{Acknowledgments}
We thank Tobias Gehring, Vitus H\"andchen, Sacha Kocsis and Christina E. Vollmer for fruitful discussions. 
This work was supported by the Deutsche Forschungsgemeinschaft (DFG), Project No. SCHN~757/4-1, by the Centre for Quantum Engineering and Space-Time Research (QUEST), and by the International Max Planck Research School for Gravitational Wave Astronomy (IMPRS-GW). J. F. acknowledges support by the Czech Science Foundation (Project No. P205/12/0577) and by the European Social Fund and MSMT under Project No. CZ.1.07/2.3.00/20.0060.

\end{document}